\documentclass[aip, reprint]{revtex4-1}
\setcitestyle{numbers,square}
\usepackage{graphicx}
\usepackage{tabularx}
\usepackage{amsmath}
\usepackage{amssymb}
\usepackage{color}
\usepackage{xfrac}
\usepackage{fancyhdr}
\usepackage[mathscr]{eucal}
\usepackage{amsfonts}
\usepackage[normalem]{ulem}
\graphicspath{{C:/Users/bkraus/Dropbox/Stuart/Publication/}}

\usepackage{etoolbox}
\makeatletter
\patchcmd{\frontmatter@RRAP@format}{(}{}{}{}
\patchcmd{\frontmatter@RRAP@format}{)}{}{}{}
\renewcommand\Dated@name{}
\makeatother

\newcommand{\vect}[1]{\boldsymbol{#1}}
\newcommand{\iotabar}{\text{$\iota\!\!$-}}
\newcommand{\change}[1]{{\color{black}#1}}

\begin{document}

\title{Theory and discretization of ideal magnetohydrodynamic equilibria with fractal pressure profiles}

\author{B. F. Kraus}
\affiliation{Department of Astrophysical Sciences, Princeton University, Princeton, NJ 08544, USA}
\email{bkraus@princeton.edu}
\author{S. R. Hudson}
\affiliation{Princeton Plasma Physics Laboratory, Princeton, NJ 08543, USA}

\date{Submitted to Phys. Plasmas 5 June 2017; accepted 12 September 2017}

\begin{abstract}
In three-dimensional ideal magnetohydrodynamics, closed flux surfaces cannot maintain both rational rotational-transform and pressure gradients, as these features together produce unphysical, infinite currents. A proposed set of equilibria nullifies these currents by flattening the pressure on sufficiently wide intervals around each rational surface. Such rational surfaces exist at every scale, which characterizes the pressure profile as self-similar and thus fractal. The pressure profile is approximated numerically by considering a finite number of rational regions, and analyzed mathematically by classifying \change{the irrational numbers that support gradients} into subsets. Applying these results to a given rotational-transform profile in cylindrical geometry, we find magnetic field and current density profiles compatible with the fractal pressure. 
\end{abstract}

\maketitle

\section{Introduction}

Many models applied to plasmas exhibit fractal behavior \cite{Aguirre}. Energy cascades across many self-similar scales of eddies in turbulence theory \cite{Escande, Mathias}, and fractal constructs like the Hausdorff dimension have characterized properties of turbulence experimentally \cite{Budaev}. Magnetic field lines are integrable in axisymmetric configurations, but three-dimensional (3D) effects break up the paths into magnetic islands and fractally chaotic volumes \cite{Meiss}. Applied to the structure of toroidal field lines, the Kolmogorov-Arnol'd-Moser (KAM) theorem \cite{Arnold} explains the perseverance of particularly irrational flux surfaces, which remain closed despite integrability-destroying perturbations to the boundary. In this work, we investigate how fractality appears in a simple model of plasmas, ideal magnetohydrodynamics (MHD).

A frequent first approach to finding plasma equilibria, ideal MHD treats the plasma as a perfectly conducting fluid. Equilibria exist if macroscopic forces from the pressure $p$, current density $\vect{J}$, and the magnetic field $\vect{B}$ are in balance, as described by
\begin{equation} \nabla p = \vect{J} \times \vect{B}.\end{equation}
Solutions to this force-balance equation are straightforward for axisymmetric systems \cite{SolovevShafranov}, but incorporating 3D effects can introduce fractal properties into $p, \vect{B}$, and $\vect{J}$. 

In this work, we seek strictly ideal, zero-Larmor-radius MHD equilibria that satisfy three conditions. First, the pressure must be non-zero somewhere, such that a pressure profile peaked on-axis satisfies \mbox{$p(0) \neq p(r \to \infty)$}. Second, the current \mbox{$\int_C \vect{J}(\vect{r}) \cdot d\vect{A}$} must be finite for any surface $C$ and its surface element $d\vect{A}$. Third, the magnetic field $\vect{B}$ should be continuous. Several sets of ideal equilibria satisfy these conditions concurrently, but all of them rely on either non-smooth features or quantities that are discretely defined \cite{Hudson2017}.

A smooth profile is defined to be continuously differentiable, namely $C^1$. Conversely, we define a non-smooth profile to be $C_0$, so it is continuous itself but has discontinuities in its first derivative. Even though equilibria with nested flux surfaces often assume smooth pressure and current profiles, resonant perturbations to plasma boundary conditions can lead to unphysical infinite currents where the pressure smoothly varies. This problem can be avoided by allowing for non-smooth features and fractality. \change{As motivated below, the particular solution examined here localizes non-smooth behavior to a special subset of irrational flux surfaces.}

Infinite currents arise with smooth pressure profiles due to a classical small divisors problem \cite{Yoccoz}. We assume ideal MHD equilibria with nested flux surfaces, where $p$, $\vect{B}$ and $\vect{J}$ are all flux-surface functions. In this case, the parallel current density can be written \mbox{$\vect{J}_\parallel = \lambda \vect{B}$}, and the perpendicular current density is produced by pressure gradients: \mbox{$\vect{J}_\perp = \vect{B} \times \nabla p/B^2$}. Enforcing charge conservation \mbox{$\nabla \cdot \vect{J} = 0$} gives \mbox{$\nabla \cdot (\lambda \vect{B}) = \vect{B}\cdot \nabla \lambda = - \nabla \cdot \vect{J}_\perp$}. Assuming toroidal geometry with straight-field-line coordinates, the Fourier components of the differential operator $\vect{B}\cdot \nabla$ become \mbox{$(\vect{B}\cdot\nabla)_{mn} = i\sqrt{g}^{-1}(m \iotabar - n)$}, where $\iotabar$ is the rotational-transform and $m,\ n$ are the poloidal and toroidal mode numbers, respectively \cite{SolovevShafranov}.  Solving for each harmonic of the parallel current's Fourier coefficient gives  
\begin{equation} 
\label{eqn:generalB} \lambda_{mn} = \frac{i(  \sqrt{g}  \nabla \cdot \vect{J}_\perp)_{mn}}{m\iotabar - n}  + \Delta_{mn} \delta (\iotabar - n/m), 
\end{equation}
where $\sqrt{g}$ is a Jacobian that satisfies \mbox{$\sqrt{g}^{-1} = \vect{B} \cdot \nabla \zeta$} for the toroidal coordinate $\zeta$.

Across resonant flux surfaces $\iotabar = n/m$, the $\delta$-function spike in current density is integrable and only produces current discontinuities. However, the first term shows the current density behaves like $1/x$ near a resonant flux surface at $x = 0$; integrating such a current density leads to a logarithmically divergent current. To eliminate these unphysical infinite currents at resonances, either term in the numerator must be zero where $\iotabar$ is rational. The Jacobian $\sqrt{g}_{mn}$ depends on the field geometry, so it does not vanish for arbitrary boundary conditions. Therefore, a 3D equilibrium must satisfy $\nabla p = 0$ on all resonances $\iotabar = n/m$ to globally avoid infinite currents.

Plasmas in models other than ideal MHD remedy these resonant currents by changing field topology. In resistive MHD, tearing modes open resonant rational surfaces into islands, wherein the pressure is uniform to zeroth order. When field line structure is taken as a Hamiltonian system, overlapping islands lead to chaotic field lines that fill volumes ergodically \cite{Chirikov} so that the plasma pressure becomes uniform within each chaotic region. In contrast, ideal plasmas obey Alfv\'{e}n's frozen flux theorem so that field topology is fixed. Equilibria that begin with nested flux surfaces must keep them for all time if variations are ideal. Thus, the perturbed system must accommodate infinite currents or else tolerate a non-smooth pressure profile. 

If the pressure is flat near every rational, but not uniformly trivial ($p(r) = 0$), then a fractal profile must emerge where the pressure \change{is only allowed to change} on flux surfaces with irrational rotational-transform. \change{Crucially, many irrational numbers are so near to neighboring rationals that they are enveloped by flattened rational regions. It is a central task of this work to examine which subset of the irrational surfaces can maintain pressure gradients, and whether this subset is large enough to have physical significance.}

We restrict our attention to ideal MHD to focus on only one of two effects that complicate 3D equilibria. Plasmas in toroidal geometry exhibit (i) stochastic field lines that wander chaotically about the plasma volume, and (ii) a corresponding flattening of the pressure profile which ensures \mbox{$\vect{B} \cdot \nabla p = 0$}. Though the two effects arise together, the second can be studied alone in a simplified plasma model that is both ideal and two-dimensional. How can a fractal pressure be discretized appropriately for equilibrium calculations, and how will the fractal structure be reflected in other plasma parameters? These questions can be answered by considering ideal MHD equilibria in cylindrical geometry with fractal pressure as an input, allowing the effects of fractality to be studied in isolation.

The physics of this fractally flat pressure profile will be examined in this paper. In Section \ref{section:Diophantine}, a method of generating a physical pressure profile will be shown, grounded in the Diophantine condition of the KAM theorem; this pressure profile will be shown to be fractal. Section \ref{section:grid} describes various algorithms for approximating the fractal pressure profile numerically, and Section \ref{section:robust} employs number theory to sort the irrational numbers by distance from the rationals. Section \ref{section:profiles} demonstrates ideal MHD equilibria compatible with the discretized fractal pressure, as well as smoothed equilibria that approximate the resistive case.

\section{\label{section:Diophantine}Rational intervals and the Diophantine pressure}

Ideal MHD cannot tolerate pressure gradients on resonant rational surfaces. This constraint applies not just to the rational number itself, but a neighborhood of finite extent that surrounds it. The small-denominator singularity $1/(m \iotabar - n)$ extends to nearby locations, since $\int_\epsilon^\zeta dx / x$ is unbounded for $\epsilon < \zeta \ll 1$ as $\epsilon \to 0$. Thus, an entire neighorhood $(\frac{n}{m} - \delta, \frac{n}{m} + \delta )$ must be flattened around each rational $\iotabar$ for some choice of $\delta = \delta(m,n)$.

This situation of defining neighborhoods around rational centers is reminiscent of the Diophantine condition in KAM theory, a concept which bounds the effective width around each rational that is affected by resonance.  Perturbations at rational $\iotabar$ become more severe as the system deviates further from axisymmetry, so the Diophantine neighborhoods around each rational increase in width with a larger perturbation magnitude. Furthermore, the order of a given rational surface also matters: higher-order rationals (large $m$) have less importance and affect a smaller neighborhood than low-order rationals (small $m$). The KAM theory posits a width $\delta$ that satisfies these conditions: $\delta(m,n) = d / m^k$, where $d$ and $k$ are parameters that can be tuned to approximate the shape of a general 3D perturbation. 

\subsection{Defining the Diophantine set and pressure}

In the spirit of flattening the pressure profile where the rotational-transform is too nearly rational, we define the rational set $\iotabar_R = \iotabar_R(d,k)$ as
\begin{equation} 
\iotabar_R(d,k) = \bigg\{ \iotabar: \iotabar \in \bigcup_{\forall n, m} \left( \frac{n}{m} - \frac{d}{m^k}, \frac{n}{m} + \frac{d}{m^k} \right)
\bigg\}.
\end{equation}
In other words, $\iotabar$ is ``effectively" rational if it lies within a distance $d/m^k$ of a rational number. A flattened region surrounds each rational number, though these regions can overlap and merge if their widths $2d/m^k$ are large enough.

Even among this infinite number of rational intervals, some irrational numbers fall outside of the set $\iotabar_R$. This class forms the sufficiently irrational Diophantine set $\mathcal{D}_{d,k}$, defined as the complement of the rational set: $\mathcal{D}_{d,k} \equiv \mathbb{R} \setminus \iotabar_R(d,k)$. The more general Diophantine set $\mathcal{D}$ is the union of all $\mathcal{D}_{d,k}$ for any $d$ and $k$. 

The set of Diophantine numbers $\mathcal{D}_{d,k}$ for a particular $d$ and $k$ is a subset of the irrationals. Since every irrational number is infinitesimally close to a rational---ie., the irrationals are accumulation points for the rationals---every Diophantine point directly neighbors a rational neighborhood. Thus, all reals in the Diophantine set are totally disconnected from one another.

The two Diophantine parameters, namely $d$ and $k$, strongly affect the distribution of Diophantine numbers and the shape of the resulting pressure profile. The magnitude parameter $d$ measures the strength of a 3D perturbation: as $d$ increases, the size of every rational neighborhood increases linearly. The scaling parameter $k$ determines the decay of this perturbation at higher-order resonances. A higher value of $k$ causes rational windows to shrink rapidly at high-order resonances, as in the case where several large islands dominate with many nested flux surfaces in between; a lower $k$ allows a perturbation to penetrate down to higher-order rational resonances, and is analagous to equilibria with many small islands nested between each other.

\begin{figure}
\centering
\includegraphics[width=0.5\textwidth]{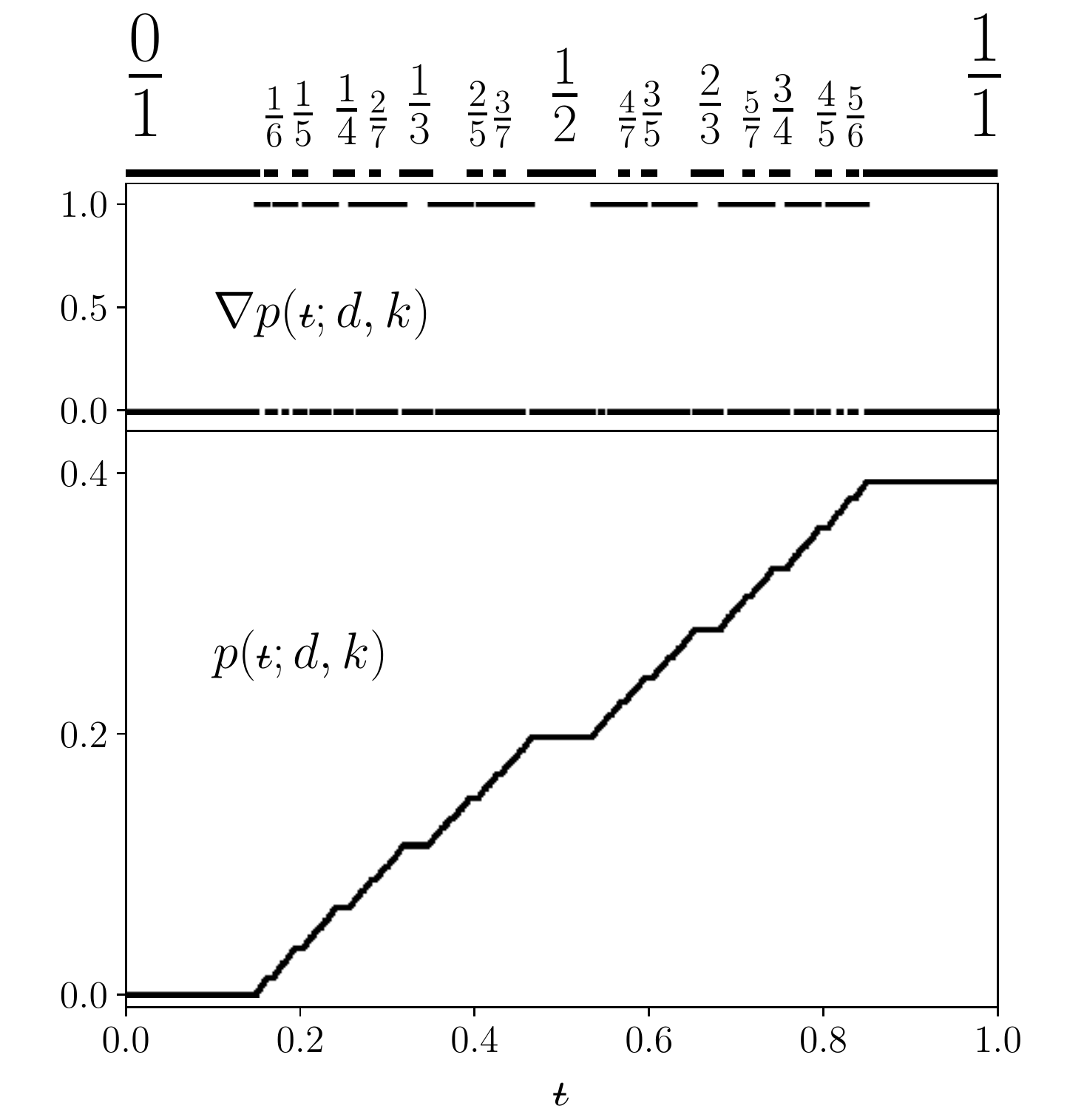}
\caption{\label{fig:gradPandP}An illustration of low-order rational windows (top), the resulting Diophantine pressure gradient (middle), and its integral (below), all shown versus rotational-transform $\iotabar$. The above calculations used Diophantine parameters $d = 0.15$ and $k = 2.1$. For legibility, the profile is only flattened on a small number of rationals, with five levels of the Farey tree shown ($n/m \in \mathcal{F}_5$, defined in Section \ref{section:grid}).}
\end{figure}

We choose to examine a pressure gradient which is nonzero only where the rotational-transform is Diophantine:
\begin{equation} \frac{dp(\iotabar; d, k)}{d\iotabar} = \begin{cases} 1, & \iotabar \in \mathcal{D}_{d,k}; \\ 0, & \textup{otherwise.}\end{cases} \end{equation}
In this case, the pressure gradient is pointwise discontinuous on all Diophantine irrationals and trivial otherwise. A discretized example profile $p(\iotabar)$ is shown for $d = 0.15$ and $k=2.1$ in Fig.~\ref{fig:gradPandP}.

For decades, non-smooth pressure profiles have been regarded as unconventional and even disastrous. Grad \cite{Grad} anticipated that staircase pressure profiles would arise in ideal, toroidally confined systems, terming the resulting fractal system ``pathological." Despite this early judgment, the non-standard Diophantine profile is tractable in the light of modern measure theory and knowledge of fractal structures \cite{ELee, Mandelbrot}.

\subsection{Non-emptiness and topology of the Diophantine set}

Before examining the physics of the Diophantine pressure, we must ensure the mathematical possibility of this construction. Since all Diophantine points are disconnected, there is no interval of any length where the pressure has non-zero gradient throughout. How can an overall pressure gradient exist where only discrete points and not intervals contribute to changes in pressure? Strictly speaking, the only possibility is that the Diophantine set is uncountably infinite for some $d$ and $k$, since it takes this cardinality and no less for a set of discrete points to span non-zero measure. 

While number and measure theory are necessary for further understanding of this issue (see Section \ref{section:robust} for details),  one fundamental fact can be derived from surprisingly simple arguments: a non-trivial fractal pressure exists for some $d$ and $k$. Proving this fact about the Diophantine set relies on solving the problem in reverse, or finding an upper bound on the measure of its complement. In a method akin to the proving the vanishing measure of the Cantor set, we consider the total size of rational regions. Precise measurement of this size requires checking for overlaps between each pair of rational regions, a process which is computationally expensive but which only reduces the total width of rational regions. Ignoring overlap thus \emph{overestimates} the measure of $\iotabar_R$ and \emph{underestimates} the measure of irrational points in $\mathcal{D}_{d, k}$. As such, we find an upper bound on the total rational length by evaluating the sum:
\begin{equation} \text{Rational length } L = \sum_{m=1}^\infty \frac{2d}{m^k} \phi(m),\end{equation}
where $\phi(m)$ is Euler's totient function \cite{Lehmer}. Since $\phi(m)$ counts all integers less than $m$ which are relatively prime with $m$, it is always true that $\phi(m) < m$. Therefore, \mbox{$L < 2d \sum_{m=1}^\infty m^{-(k-1)}$}. This expression is the well-known hyperharmonic series, which converges if \mbox{$(k-1) > 1$} and diverges otherwise. As long as $k > 2$ and $d$ is sufficiently small, the length $L$ has a value less than one, and there is space left over for the Diophantine numbers. Since the measure of the Diophantine numbers is non-zero for some choices of $d$ and $k$, there exists a non-trivial Diophantine pressure profile. \change{In other words, the resulting equilibrium can indeed confine a plasma, even if flattened regions lie around every rational flux surface.}

A pivotal result from this proof concerns the critical case of $k = 2$. At this point, the rational regions exactly cover all space, and the pressure becomes uniformly zero. Still, a countable infinity of irrationals exist in the Diophantine set at $k = 2$, \change{nested between the flattened pressure regions}. This understood boundary case will be mentioned frequently in the following.

Topologically, the Diophantine pressure profile is similar to the fat Cantor set \cite{Umberger}, which is also totally disconnected and can have either empty, partial, or full measure. Other research has benefited from such comparisons, for instance by providing fat-fractal scaling laws for islands around KAM surfaces \cite{Hanson}. Here, we suggest that the set of Diophantine numbers is homeomorphic to the fat Cantor set for some $d$, a connection suggested by Brouwer's theorem \cite{Brouwer}. The set of Diophantine numbers $\mathcal{D}_{d,k}$ is already much like the Cantor set, in that it is (i) non-empty for some small $d >0 $ and $k \ge 2$, (ii) compact (closed and bounded), (iii) totally disconnected, and (iv) metrizable (which is true of any set in the real numbers). 

However, to definitively satisfy Brouwer's theorem, the Diophantine numbers must also be a perfect set, which by definition contains no isolated points. For the Diophantine numbers to comprise a perfect set, every neighborhood $(\iotabar - \epsilon, \iotabar+\epsilon), \epsilon > 0$ around elements $\iotabar \in \mathcal{D}_{d,k}$ must contain another Diophantine number. While this is true of some elements of the Diophantine set, we have not been able to prove that it is true for all elements $\iotabar$. Further, there are certain values of $d$ where the set is clearly \emph{not} perfect, such as the highest $d$ before $\mathcal{D}_{d,k}$ is empty: here, only two well-separated noble numbers remain in the set. Nonetheless, it seems likely that there are particular values of $d$ and $k$ for which the Diophantine set can be understood as a Cantor-like distribution of irrationals.

\subsection{Self-similarity of mediant windows}

In densely packed regions of Diophantine irrationals, the distribution of numbers can be shown to be fractal. With this insight comes a series of techniques for analyzing self-similar functions, parallels which should be exploited to gain further insight into implications for ideal MHD equilibria.

Fractal objects are self-similar on all scales \cite{Mandelbrot}. A literal definition of self-similarity can be applied to discrete systems like the distinct rational intervals considered in the Diophantine set. In this case, the relative sizes of rational and irrational regions at a given scale can be a proxy for length, so a constant ratio of sizes over several levels of rational perturbations is indicative of self-similarity and fractality.

As will be explained in more detail in Section \ref{section:grid}, the rational regions in a given interval can be built up from low- to high-order by selecting the mediant rationals between two neighboring rationals $p_0/q_0$ and $p_1/q_1$. The first mediant of these two rationals, guaranteed to lie between them and higher order such that $q_2 > q_0, q_1$, is $p_2 / q_2 = (p_0 + p_1)/(q_0 + q_1)$. The next mediant, between $p_1/q_1$ and $p_2/q_2$, will be higher order still. We will study neighboring pairs of rationals in these mediant sequences.

\begin{figure}
\centering
\includegraphics[width=0.5\textwidth]{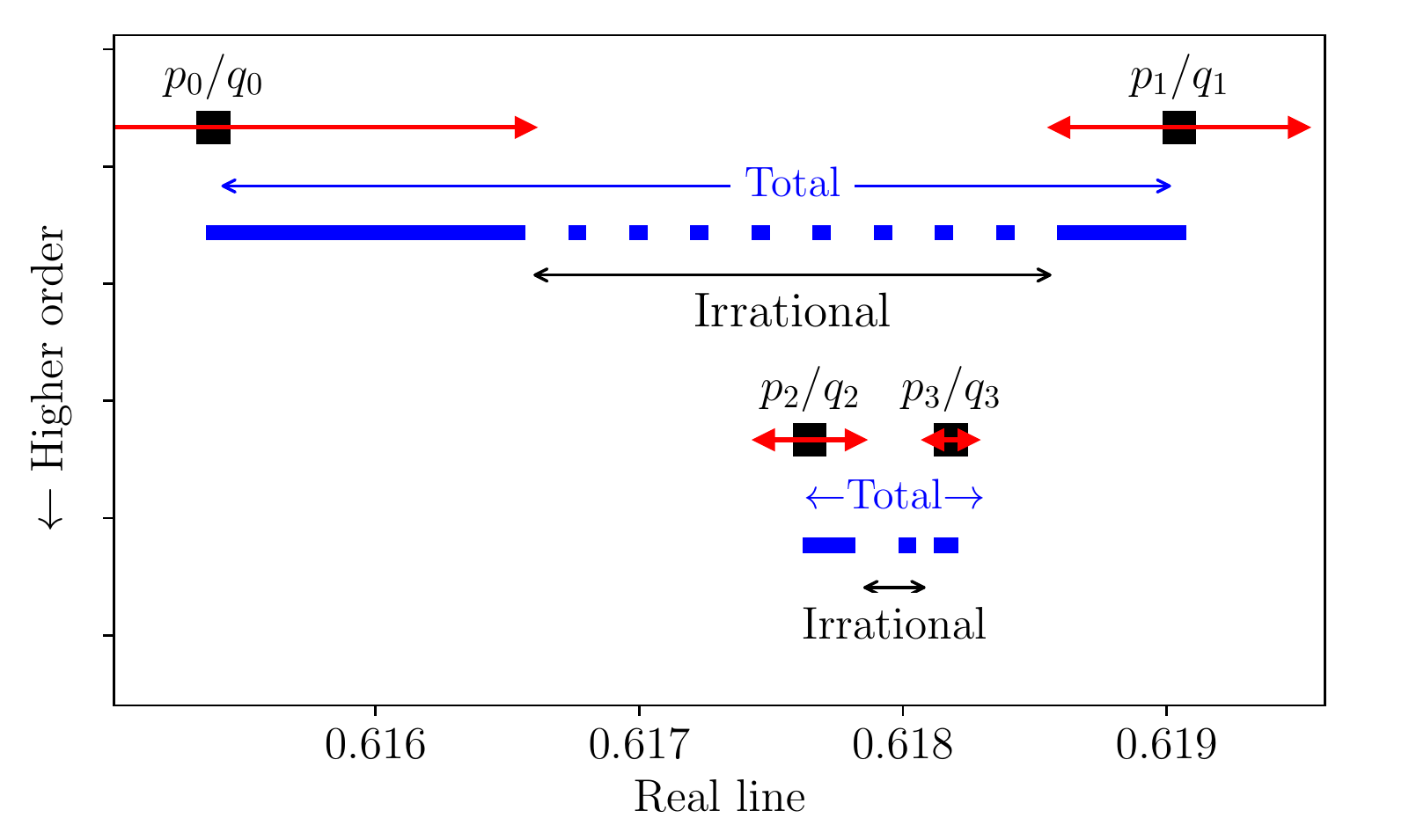}
\caption{\label{fig:fractalWindows}Two pairs of mediant rationals are shown, $\{ p_0/q_0, p_1/q_1\}$ and $\{ p_2/q_2, p_3/q_3\}$. The rational windows around each rational are shown in red, and the Diophantine irrational region is shown in black in between. If the ratio of lengths (Irrational/Total) is constant for pairs of mediants at all scales, the distribution of rational windows is fractal.}
\end{figure}

If the profile scales self-similarly, the irrational region between two neighboring rationals should have constant length in proportion to the distance between the rationals. In other words, the ratio (Irrational/Total) (shown in Fig.~\ref{fig:fractalWindows}) must be constant for rationals of all orders. We begin with two rationals $p_0/q_0$ and $p_1/q_1$, and define the irrational interval between them as
\begin{equation} \bigg( \frac{p_0}{q_0} + \frac{d}{q_0^k}, \frac{p_1}{q_1} - \frac{d}{q_1^k} \bigg). \end{equation}
Continuing to larger denominator rationals on an alternating path, the next mediants are
\begin{equation} \frac{p_2}{q_2} = \frac{p_0 + p_1}{q_0 + q_1}, \qquad \frac{p_3}{q_3} = \frac{p_0 + 2p_1}{q_0 + 2q_1},\end{equation}
and the corresponding irrational interval lies between \mbox{$p_2/q_2 + d/q_2^k$} and \mbox{$p_3/q_3-d/q_3^k$}.

If these two levels are to be self-similar, the ratio of the irrational length to the total length must be constant:
\begin{equation} \frac{\displaystyle \bigg(\frac{p_1}{q_1} - \frac{d}{q_1^k} \bigg) - \bigg(\frac{p_0}{q_0} + \frac{d}{q_0^k} \bigg)}{\displaystyle \bigg(\frac{p_1}{q_1} - \frac{p_0}{q_0}\bigg)} = \frac{\displaystyle\bigg(\frac{p_3}{q_3} - \frac{d}{q_3^k}\bigg) - \bigg(\frac{p_2}{q_2} + \frac{d}{q_2^k}\bigg)}{\displaystyle\bigg(\frac{p_3}{q_3} - \frac{p_2}{q_2}\bigg)}.\end{equation}
Using the definitions for $p_2, p_3, q_2$, and $q_3$ implied above eventually gives:
\begin{equation} \frac{q_1^k + q_0^k}{(q_0 + 2q_1)^k + (q_0 + q_1)^k} = \bigg[ \frac{q_0 q_1}{q_0^2 + 3q_0q_1 + 2q_1^2} \bigg]^{k-1}.\end{equation}

The case of $k = 2$ has been shown to be a boundary case for the Diophantine numbers. Assuming $k=2$ in the above leads to an analytically solvable equation with the following meaningful solutions:
\begin{equation} q_1 = \pm \frac{\sqrt{2}}{2} q_0 \quad \text{and} \quad q_1 = \frac{1 \pm \sqrt{5}}{2} q_0,\end{equation}
where $(1+\sqrt{5})/2 = \varphi$ is the golden mean. Any possible ratio $q_1 / q_0$ that fulfills the fractal condition must be irrational, which is strictly impossible given that $q_0, q_1$ are integers by definition.

However, it is possible that $q_{i+1} / q_i \to \varphi$ for large values of $i$, or for relatively high-order rational resonances. Is there a number whose series of converging mediants has denominators $q_0, q_1, q_2\dots$ such that the eventual ratio of successive denominators approaches the golden mean $\varphi$?

A relevant limit, first noted in 1611 by Johannes Kepler \cite{Dunlap}, shows that any Fibonacci sequence $\{F_i\}$ satisfies this property at large $i$:
\begin{equation} \lim_{i \to \infty} \frac{F_{i+1}}{F_i} = \varphi.\end{equation}
It is well known that the sequence of mediants $\{p_i/q_i\}$ that converge to the golden mean itself have denominators $q_i$ that form a Fibonacci sequence, and thus $\lim_{i \to \infty} q_{i+1}/q_i = \varphi$. More generally, the sequence of mediants that converges to \emph{any} noble number has the same large-$i$ limit as above. Thus, the fractal pressure is perfectly self-similar at $k = 2$ in the regions near each noble number. \change{As the nobles form a dense subset of the real numbers, this restriction does not imply that fractality is an intermittent property of the Diophantine set.} In this sense, the Diophantine pressure profile is demonstrably fractal. 

This result was shown for due to its analytic tractability, but it applies to a trivial result since the Diophantine pressure is uniformly zero at \mbox{$k = 2$}. However, keeping $d$ constant and increasing $k$ only expands the Diophantine set, so more irrationals become elements and none are lost. It is thus feasible that the Diophantine numbers maintain their self-similar character for higher $k$ where the pressure profile is non-trivial.

\section{\label{section:grid}Discretization with a fractal grid}

This section will explain which numerical representations of the Diophantine pressure profile successfully converge to the fractal profile as grid resolution increases. 

For purposes of comparison, it is necessary to define a proxy for ``convergence.'' The fractal profile can never be calculated exactly, so a computer cannot directly measure the difference between the exact version and an approximation. However, a well-defined quantity can be used despite the endless intricacies of the fractal pressure: the integrated pressure in the system, 
\begin{equation}
p_\text{max} \equiv \int p'(\iotabar) d\iotabar \approx \lim_{N \to \infty} \sum_{i=1}^N p'(\iotabar_i) \Delta \iotabar_i,
\end{equation}
where $\iotabar$ is taken as a proxy radial coordinate. Note that the approximate equality is valid for Riemann-integrable discretized functions, but is never valid for the full Diophantine pressure. If a discretized profile matches $p_\text{max}$ to the exact value from the fractal case, the total size of the gradient-sustaining regions (consisting of irrational points in the fractal case and ``irrational'' intervals in the discretized version) is well approximated. For instance, it has been shown that for $k=2$ no gradients remain in the fractal pressure so that $p_\text{max}(k = 2)$ ought to tend toward zero. Though this instance leads to a trivial pressure, it is chosen as a test case here because the exact solution is known.

Let $\{ \iotabar_i \}_{i=0}^N$ be a finite set of grid points used to model $p'(\iotabar)$. Out of convenience, most numerical methods use regularly spaced grids, where $\Delta_i \equiv \iotabar_i - \iotabar_{i-1}$ is a constant for all $i$. The most trivial example cases define a normalized domain, spanning from 0 to 1, such that the values $\iotabar_i$ lie on rational numbers $i/N$. Due to the definition of the Diophantine $p'(\iotabar)$, all values $\iotabar_i$ lie within flattened rational regions: $p'(\iotabar_i) = 0$ for all $i$. For the $k=2$ case, this answer is correct, but such a scheme can never predict the shape of $p(\iotabar)$ for higher $k$. The method neglects all locations where the pressure changes. 

A straightforward alternative strategy redefines the domain: let $\iotabar$ span from zero to some irrational value $\xi$, which sets the grid points $\iotabar_i = i \xi/N$ to be irrational. Depending on the precise values of these chosen grid points, some $\iotabar_i$ may lie outside of all rational windows such that $p'(\iotabar_i) = 1$. As the number of grid points $N$ increases, more irrationals will join the pool of selected quantities so the approximation converges toward a certain shape for $p(\iotabar)$.

\begin{figure}
\centering
\includegraphics[width=0.5\textwidth, bb=0 0 405 445]{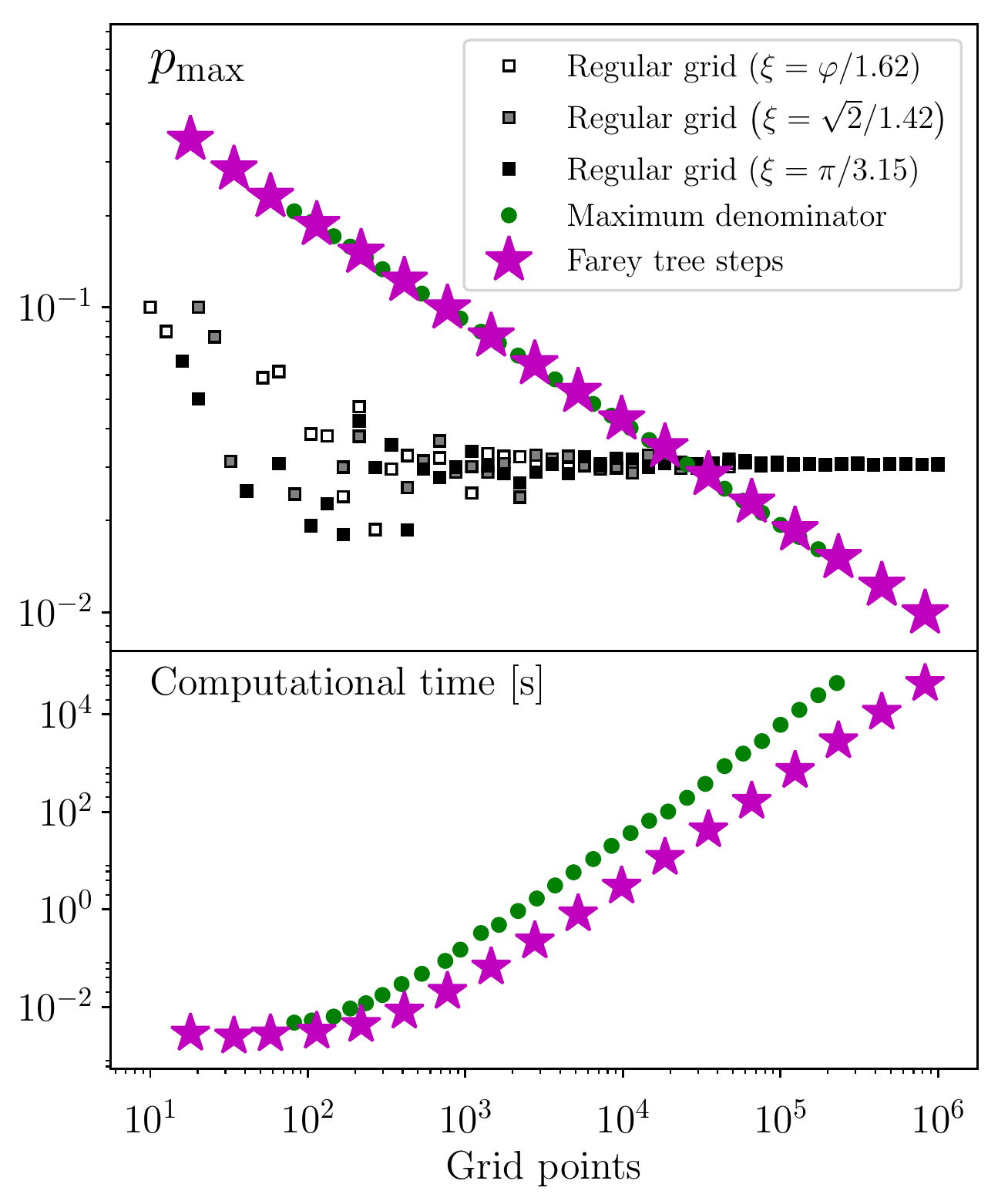}
\caption{\label{fig:convergence}A comparison of different discretization methods for a known case, $k = 2$, where $p_\text{max} = 0$. The top figure plots $p_\text{max}$ as a function of grid resolution, showing power law convergence for irregular grids based on the Diophantine condition and inaccurate convergence for regular grids on irrational values. The bottom figure compares computational time for the two methods on irregular grids.}
\end{figure}

Nonetheless, this scheme of selecting regularly spaced irrational numbers is not aligned with the pressure profile under consideration. For instance, it was shown that the $k=2$ Diophantine pressure is fractal near noble values of $\iotabar$, which means an infinite number of Diophantine points can exist near each noble number if $d$ is small enough. A regular grid of irrationals may choose one of these points, or it may land within a rational region and not contribute: in any event, the correspondence between the grid and the gradient-supporting Diophantine values is essentially random. Increasing the number of grid points to any higher but still finite value will not correct this situation, meaning the numerical scheme does not converge to the correct answer. Fig.~\ref{fig:convergence} shows this errant behavior for several regular grids with different irrational endpoints $\xi \approx 1$ (shown as white, grey, and black squares). Note that these approximations asymptote to the incorrect value of $p_\text{max}$, and are thus informed not by the true fractal profile (which has $p_\text{max} \to 0$) but by numerical details of where grid points lie.

More satisfying convergence properties can be attained by incorporating the Diophantine set directly into the procedure for building the grid. Distributing grid points at the edges of each rational interval leads to a well-behaved numerical scheme, so the error reliably decreases with increasing resolution. Since an infinite number of flattened regions cover much of the real line, a discretization of this grid must choose the most important subset of these regions that allows computational efficiency without sacrificing mathematical accuracy. In effect, this requirement necessitates an ordering of the rational numbers. The ordering should prioritize the rationals with small denominators, but also needs to select values which are relatively far from other rationals that have already been chosen.

\begin{figure}
\centering
\includegraphics[width=0.5\textwidth]{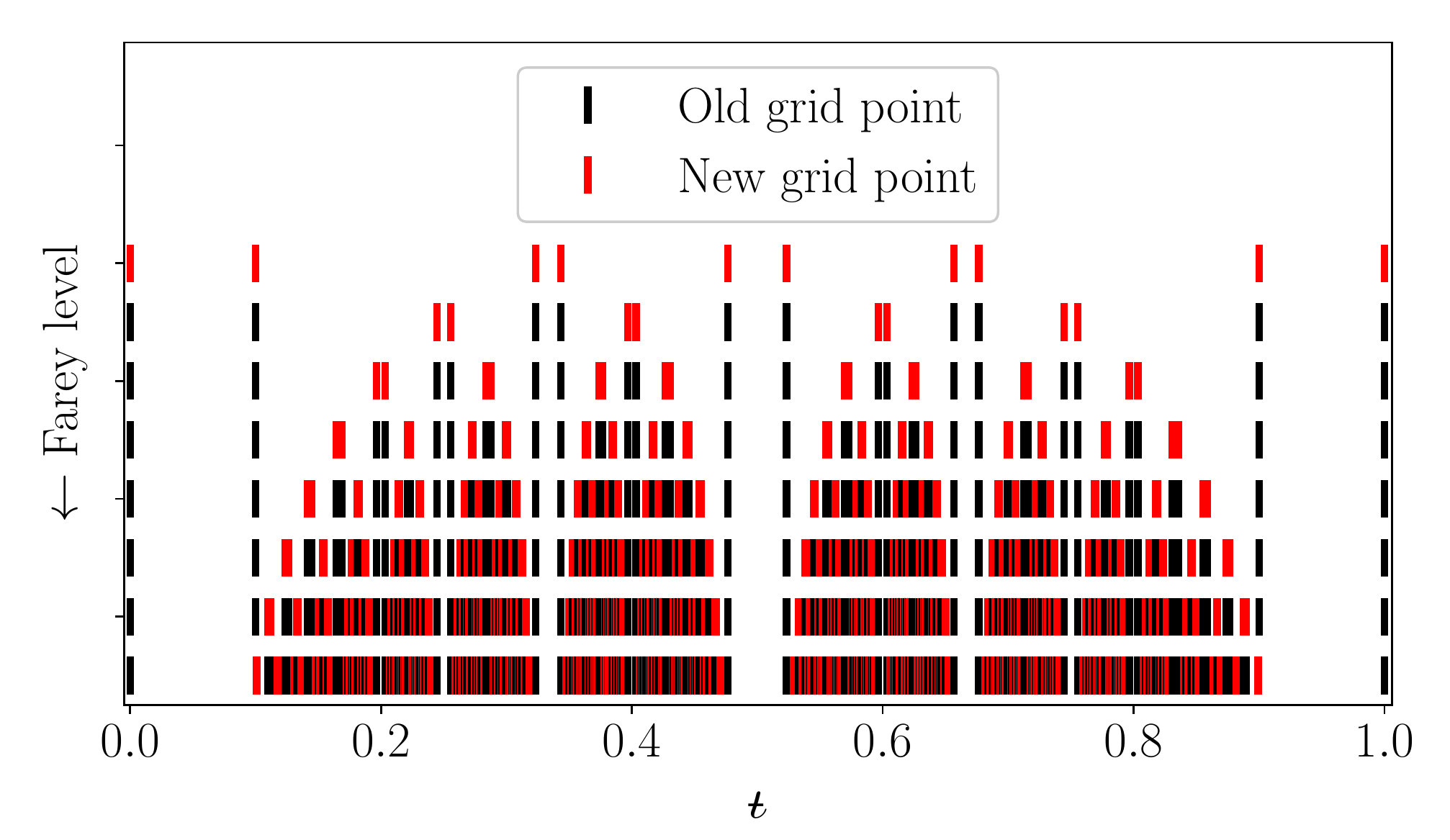}
\caption{\label{fig:gridpoints}A series of fractal grids, shown from low resolution (top) to high resolution (bottom) and calculated with  $d = 0.1$ and $k = 2.1$. Each grid contains the lower-resolution grids above it (black), but has added new points (red) that border the rational windows of higher-level Farey trees.}
\end{figure}

Such a scheme is realized by Farey trees, which begin with two neighboring rationals and successively list every higher-order rational in between. If $\iotabar(r) \in [0, 1]$, then an $L$th-order truncated Farey tree $\mathcal{F}_L$ can be derived from the initial condition $\mathcal{F}_0 = \{ 0, 1 \}$ recursively:
\begin{equation} \begin{split} \mathcal{F}_i^* &= \bigg\{ \frac{n_L + n_R}{m_L + m_R} \bigg|\ \forall \text{ adjacent } \frac{n_L}{m_L}, \frac{n_R}{m_R} \in \mathcal{F}_{i-1} \bigg\};  \\ 
\mathcal{F}_i &= \mathcal{F}_i^* \bigcup \mathcal{F}_{i-1}. \end{split} \end{equation}
The points of interest for the Diophantine pressure profile are not the rational numbers in $\mathcal{F}_L$,
but the edges of the flattened regions around them: \mbox{$\{ n/m \pm d/m^k \ |\ \forall n/m \in \mathcal{F}_L \}$}.

As more rationals are added with each Farey tree order, many flattened regions will overlap other, lower-order rationals that have already been considered. Each new rational region \mbox{$(n/m - d/m^k, n/m + d/m^k)$} must be compared to other nearby regions. If the interval lies entirely inside a larger interval, it is ignored; if it lies partially inside another interval, the lower-order interval is redefined to encompass a wider flattened region.

\begin{figure}
\centering
\includegraphics[width=0.5\textwidth, bb=0 0 425 183]{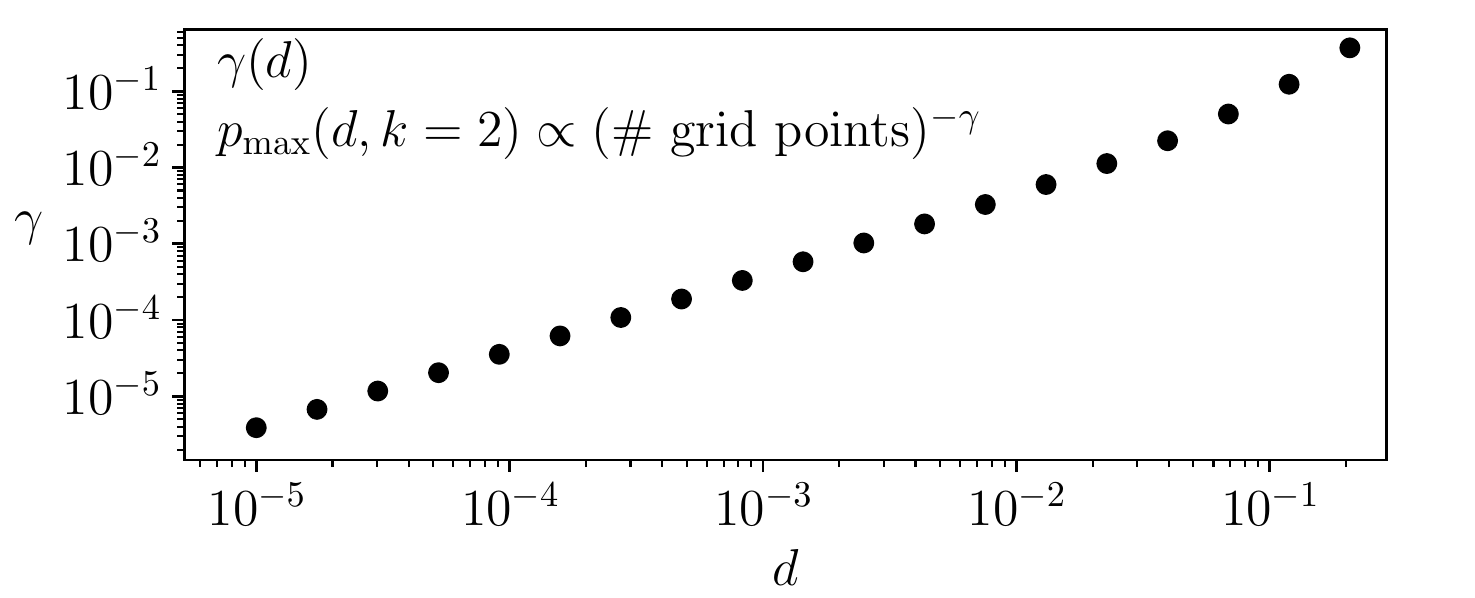}
\caption{\label{fig:powerLawScalingD}The rate of convergence of the fractal grid for $k = 2$.
Higher values of $d$ converge much more rapidly with fewer grid points.}
\end{figure}
    
Thus, the scheme of approximating a fractal grid is as follows. Choosing the Diophantine parameters $d$ and $k$ from the start, the method calculates $\{ n/m \pm d/m^k\}$ for all rationals in $\mathcal{F}_0$. This scheme repeats for each $\mathcal{F}_i$ from $\mathcal{F}_1$ to the desired resolution $\mathcal{F}_L$, with the addition of a check for overlaps with previous regions. Fig.~\ref{fig:gridpoints} displays several possible resolutions of such grids. 

Both the regular- and irregular-grid methods of discretization are applied to the known test case \mbox{$k = 2,\ p_\text{max} \to 0$}, with results shown in Fig.~\ref{fig:convergence}. Unlike regularly spaced grids, the Farey tree grid has desirable convergence properties: $p_\text{max}$ follows a power law toward the infinite-resolution limit. Furthermore, the Farey tree is also a relatively efficient method for ordering the rationals. For example, consider an alternate scheme that develops a grid with all rationals $n/m$ with \mbox{$m < M$}, for some finite integer $M$. This process, labeled ``Maximum denominator'' in Fig.~\ref{fig:convergence}, converges in the same way that the Farey tree grid does; however, it counts rationals in an inefficient order and wastes more time discarding overlapping regions. This approach takes five times longer than using the Farey tree method for resolutions that have comparable numbers of grid points. 

Interestingly, the power law convergence of $p_\text{max}$ to its ultimate value with increasing numerical resolution is dependent on the Diophantine parameters. For $k = 2$, the scaling is shown versus $d$ in Figure \ref{fig:powerLawScalingD}. If the maximum pressure scales as $p_\text{max} \propto N^{-\gamma}$, then the figure shows $\gamma$ change as a function of $d$. For small $d$, tiny rational windows are flattened around each rational, so the pressure takes many more rationals to converge to a reasonable approximation; large $d$ have much larger $\gamma$ and thus converge more quickly to accurate $p_\text{max}$. A power-law scaling for $\gamma(d)$ is evident for $d \ll 0.1$; as $d$ increases, though, this relationship is distorted due to overlapping of rational regions.

\change{
\section{\label{section:robust}Classification and Robustness of Diophantine Irrationals}}

Discretizations of the Diophantine pressure converge as resolution increases, as long as they are dealt with on fractal grids. However, whether the numerical limit accurately models the fully fractal case is a different matter. We are proposing a plasma equilibrium where the pressure only changes on sufficiently irrational flux surfaces. Where are these irrationals distributed in space, and how much plasma volume do they cover? \change{The following scrutiny of this question is critical for full understanding of the fractal system: the answers are buried in the properties of irrational numbers, and are thus inaccessible to the discretized model described above.}

\change{The concept at hand is measure}, a quantity that represents the size of a set of numbers. In common parlance, the measure $\mu = \mu(\mathcal{A})$ is the length of the fewest possible intervals that can cover the set of numbers $\mathcal{A}$ \cite{Klambauer}. Ultimately, we seek the measure of the Diophantine set $\mu(\mathcal{D}_{d,k})$, \change{which physically represents} the fraction of the plasma volume that can support pressure gradients.

From the perspective of regular-grid Riemann integration, a paradox emerges in calculating $\mu(\mathcal{D}_{d,k})$. The Diophantine set only contains disconnected irrational numbers, yet these numbers must make up some measure $\mu(\mathcal{D}_{d,k})$ along the real line for the pressure to be anywhere non-zero. As stated in Section \ref{section:Diophantine}, the Diophantine set must be uncountably infinite for a non-trivial pressure to exist. It was shown that this must be the case because the lower bound on the measure of the Diophantine set is above zero for $k > 2$.

\change{Surpassing the specificity of this lower bound on $p_\text{max}$ is difficult}, however, since the Diophantine pressure profile is not Riemann integrable. As explained in Section \ref{section:grid}, the normalized value $p_\text{max}$ does not converge in the limit $\Delta \iotabar \equiv (\iotabar_i - \iotabar_{i-1}) \to 0$. Lebesgue integration is the only viable way to count the measure of sufficiently irrational $\iotabar$. For a given $d$ and $k$, the distribution of Diophantine numbers in $\iotabar$-space can be treated as density points \cite{Bogachev}, and calculating their total measure would allow calculation of $p_\text{max}$. 

Though mathematicians such as Khinchin and Niven have made progress on the measure theory of classes of irrationals \cite{Khinchin, Niven}, we have not found a complete explanation in the literature of how $\mu(\mathcal{D}_{d,k})$ depends on $d$ and $k$. Physically, the fractal pressure profile can range from essentially unflattened (\mbox{$d \to 0$}), in which case \mbox{$p_\text{max} \to 1$}, to totally flattened (large $d$, $p_\text{max} \to 0$). Through this transition, some irrational numbers must gradually be lost from the set of gradient-supporting rotational-transform. Understanding this measure as a function of the Diophantine parameters requires knowing which numbers exist in the Diophantine set for these cases and those in between, and what fraction of the real line these numbers cover.

Immediately, \change{no rational number is Diophantine}, since every rational number is surrounded by and included inside a flattened rational window. Moreover, some portion of the irrational numbers that are too close to a rational are also discluded from the set: these insufficiently irrational numbers $\omega$ have $\left| \omega - n/m \right| < d/m^k$ for some particular $n/m$. The remaining noteworthy irrationals, relatively far from all rational intervals, should \change{have some unifying} characteristics. For reference, the following discussions on number classifications are summarized in Table \ref{table:sets}. 

\setlength{\tabcolsep}{12pt}
\renewcommand{\arraystretch}{1.7}
\begin{table*}[tp]
\normalsize
\centering
\begin{tabular}{l l l l}
  Set & Examples & Measure $\mu$ on $[0,1]$ \\ \hline
  Rationals $\mathbb{Q}$ & $\frac{1}{2}, \frac{355}{113}$ & $\mu(\mathbb{Q}) = 0$ \\
  Irrationals $\mathbb{I}$  & $\pi, \varphi = \frac{1+\sqrt{5}}{2}$ & $\mu(\mathbb{I}) = 1$ \\
  Brjuno Numbers $\mathcal{B}$ & $\varphi, [a_1, a_2, \cdots, \bar{1}], a_i \in \mathbb{N}$ 
     & $\mu(\mathcal{B}) = 1$\\
  $j$th-order Brjuno numbers \cite{ELee} $\mathcal{B}_j$
     & $[10, 10, 10, 1020^{10}, \cdots] \in \mathcal{B}_2$ &
    $\mu(\mathcal{B}_j) = 1$, but $\mu(\mathcal{B}_{j} \setminus \mathcal{B}_{j+1}) = 0$\\
  Diophantine numbers $\mathcal{D}$ & $\varphi, [1, 1, 1, M < \infty, \bar{1}]$ & $\mu(\mathcal{D}) = \mu(\mathcal{B}_\infty) = 1$ \\
  Diophantine of type $(d, k)$ $\mathcal{D}_{d,k}$ & $[1, 2, 2, \bar{1}] \in \mathcal{D}_{0.2, 2}$ & $\mu(\mathcal{D}_{d,k}) =$ function of $d, k$ \\
  Bounded-element irrationals $\mathcal{D}_{d>0, 2}$ & $[a_1, a_2, \cdots], a_i \le M$ & 
    $\mu(\mathcal{D}_{d>0, 2}) = 0$
\end{tabular}
\caption{\label{table:sets} Several classifications of real numbers, selected as foils to the Diophantine set. Each set is accompanied by an arbitrary selection of elements it contains, as well as by its measure on the real line.}
\end{table*}

Ranking irrationals \change{by their nearness to rationals} is aided by representing real numbers $\omega$ in the continued fraction form:
\begin{equation} \displaystyle \omega = \frac{\displaystyle 1}{\displaystyle a_1+\frac{\displaystyle 1}{\displaystyle a_2 + \frac{\displaystyle 1}{\displaystyle a_3 + \cdots}}} \equiv [a_1, a_2, a_3, \cdots]\end{equation}
where $\omega \in [0,1]$, and $\{a_i\}_{i=1}^N$ are all integral and are called \emph{elements} of $\omega$. Rational numbers have finite continued fraction expansions with $N < \infty$, while irrational numbers have an infinite number of elements. Truncating the continued fraction expansion gives the \emph{convergents} to $\omega$, which are closer rational approximants to $\omega$ than any other rationals with same- or lower-order denominators. \change{The size and order of these elements $a_i$ dictates the behavior of the resulting convergents, and gives direct evidence for which irrational numbers are nearest to rationals.}

For instance, \change{numbers with arbitrarily large $a_i$ are very close to their convergents, whereas others with small elements maintain a large distance from the rationals}. A definitive volume by Khinchin \cite{Khinchin} considers bounded-element irrationals whose continued fraction expansions have elements \mbox{$a_i \leq M, \forall i$}, where \mbox{$M \in \mathbb{Z}$}. Exploiting this property and other constraints on the rational approximation of irrationals, Khinchin proves that, for sufficiently small $d$, the inequality
\begin{equation} \bigg| \omega - \frac{n}{m} \bigg| < \frac{d}{m^2} \end{equation}
has no solution for integers $n$ and $m > 0$, if and only if the irrational $\omega$ has bounded elements. Due to the theorem, all such irrationals with bounded elements are Diophantine numbers for $k = 2$ and some $d > 0$, and all Diophantine numbers have bounded elements for \mbox{$k = 2$}. Increasing $k$ further only decreases the size of rational regions, so bounded-element irrationals remain Diophantine for any \mbox{$k \ge 2$}.

\change{The set of noble numbers exemplifies the bounded-element property.} For these numbers (among them the golden mean $\varphi$), \mbox{$a_i = 1$} for all $i$ greater than some integer $k$. Since noble numbers have a finite number of elements that are not one, they have a maximum element and are thus bounded-element irrationals. For this reason, they are included in the Diophantine set; in fact, they lend the pressure profile its fractal nature, as shown in Section \ref{section:Diophantine}.
  
With this relevant subset in hand, we can now investigate whether subsets like the noble numbers contribute any measure to the full set $\mathcal{D}_{d, k}$. A bounded-element irrational can be written as a countable infinity of continued-fraction elements, and each element is an integer smaller than some finite number $M$. In this way, this class of irrationals can be listed in a one-to-one correspondence with the natural numbers. As such, the bounded-element irrationals are also countably infinite, and any countable set of discrete points has no measure. Therefore, only allowing pressure gradients on the bounded-element irrationals leaves a trivial pressure profile. To have non-zero measure, the Diophantine set must include an uncountable infinity of other irrationals. 

The above line of reasoning suggests investigation of a broader classification of irrationals, \change{whose elements may approach infinity but do so at astonishingly slow rates. Such a property is found in the Brjuno sets \cite{Brjuno}.} The $j$-th order set of Brjuno numbers, $\mathcal{B}_j$, includes all irrationals $\omega$ that have a finite Brjuno sum: 
\begin{equation} \mathcal{B}_j \equiv \bigg\{ \omega : \sum_{i = 0}^{\infty} \frac{\log(m_{i+j})}{m_i} < \infty \bigg\},
\end{equation}
where $m_l$ is the denominator of the $l$-th convergent to $\omega$. The standard Brjuno set $\mathcal{B}_1 \equiv \mathcal{B}$ is uncountably infinite, so it \change{remains possible that the Brjuno numbers can support a non-zero measure}. In addition, the sets $\mathcal{B}_j$ form a strictly-increasing, nested sequence of subsets: $\mathcal{B}_{j+1} \subsetneq \mathcal{B}_j \subsetneq \mathcal{B}_{j-1} \cdots$; this stratification \change{could be likened to the} sorting of irrationals that occurs between a full-measure, low-$d$ high-$k$ Diophantine set and its zero-measure counterpart with high $d$ and low $k$.

A study by Lee \cite{ELee} has related the Diophantine numbers and the Brjuno sets with a stunning proof of inclusion: the entire Diophantine set for any $d$ and all $k \ge 2$ is contained within the infinite-order Brjuno set, $\mathcal{D}_{d,k} \subset \mathcal{B}_\infty$. In other words, the Diophantine set for $k > 2$ derives its measure from an uncountable number of irrationals whose elements are unbounded, but whose convergent denominators grow very slowly. This growth rate is slower than can be described by the Brjuno classification scheme, so comparisons between the Diophantine set and $\mathcal{B}_j$ for any finite $j$ will not lead to more understanding of $\mu(\mathcal{D}_{d, k})$. Perhaps a separate function \change{more cleanly sorts the Diophantine set of irrationals based on their nearness to rational numbers.} 

For now, a more productive line of inquiry will rely on classifying the Diophantine set via an immediately calculable parameter. As the Diophantine parameters \change{are varied and flattened rational intervals widen}, some irrational flux surfaces transition from supporting a pressure gradient to being overlapped by a nearby rational region. For a given $k$, each noble value of $\iotabar$ sees this transition occur at a particular $d$, termed $d_*$, where the closest rational interval comes infinitesimally close to the irrational surface. This critical quantity $d_*$ parameterizes the degree of irrationality for a given noble value of $\iotabar$; \change{in physical analogy, $d_*$ represents the resilience of a given irrational surface to destruction by nearby rational islands.}

The parameter $d_*(\omega, k)$ is defined for all irrationals $\omega$ by
\begin{equation} \omega \in \mathcal{D}_{d_*, k}; \quad \omega \notin \mathcal{D}_{d_* + \epsilon, k}, \epsilon > 0.\end{equation}
All numbers in the Diophantine set have $d_* > 0$. This value $d_*$ sorts flux surfaces by their ability to support a pressure gradient in a system with general nonaxisymmetric perturbations; comparing the degree of irrationality of two flux surfaces is equivalent to comparing their $d_*$. Thus, we define \emph{robustness} as a physical quantity of flux surfaces: if an irrational surface with $\iotabar_a$ has $d_*(\iotabar_a, k) > d_*(\iotabar_b, k)$ versus another surface with $\iotabar_b$,  then the surface with $\iotabar_a$ is said to be more robust than its counterpart with $\iotabar_b$.

For some demonstrative irrationals, calculating $d_*$ is numerically trivial.  Since convergents are the best rational approximants to irrationals, the parameter $d_*$ will be set by one particularly close rational $\tilde{n}/\tilde{m}$ out of all $n/m$:
\begin{equation} \bigg| \omega - \frac{\tilde{n}}{\tilde{m}} \bigg| = \min\bigg( \bigg| \omega - \frac{n}{m} \bigg| \bigg) = \frac{d_*}{\tilde{m}^k},\end{equation}
or $d_* = | \omega \tilde{m}^k - \tilde{n} \tilde{m}^{k-1} |$. Here and below, the word ``close'' does not correspond to raw distance, but instead to distance weighted by the size of denominator $m$.

As long as the closest rational $\tilde{n}/\tilde{m}$ has $\tilde{m}$ not too high, it is straightforward to calculate $d_*$. Luckily, representation by continued fraction expansion can easily inform whether the closest rational is low- or high-order. The closest rational approximant, \change{ie., the order of the rational island which most quickly overlaps an irrational flux surface,} is generally the convergent $[a_1, a_2, \cdots a_j]$, where $a_{j+1}$ is the largest element in the continued fraction expansion for the particular irrational. Thus, the closest rational is easy to find for any noble number with a relatively small number of elements greater than one, and for these numbers finding $d_*$ is straightforward. Following the same logic, any irrational $\omega$ with bounded elements $a_i$ has a well-defined $d_*(\omega, k)$.

Given this ease of computation, a database of bounded-element irrational numbers has been constructed, generated by iterating over continued fraction representations. These values can be sorted according to their values of $d_*$, in search of number-theoretical rules that govern the robustness of these surfaces in relation to one another. Strict patterns are not easy to come by, with some large-scale trends breaking down for outlier pairs of irrationals. Numbers that otherwise seem nearby can have drastically different $d_*$, based on properties not immediately discernible through continued fraction analysis or classification into sets.

This task of seeking the most irrational flux surfaces is heavily reminiscent of an earlier search for stochastic transitions in the standard map. Greene calculated a quantity $f$, the mean residue, through laborious iteration of the map, and used it to classify surfaces by the irrationality of their rotational-transform \cite{Greene1979}. The Diophantine quantity $d_*$ is much easier to calculate than Greene's residue $f$. However, the two quantities can be shown to mirror each other, down to the numerical details of Greene's conclusions. Both methods yield sortings of the irrationals that are intimidating to analyze with pure theory, but which obey well-defined rules that can be demonstrated by extensive numerical experimentation. Below, we parallel each of Greene's assertions about the residue $f$ with directly analogous statements that can be made about the Diophantine parameter $d_*$:

\begin{figure}
\centering
\includegraphics[width=0.5\textwidth, bb=0 0 525 300]{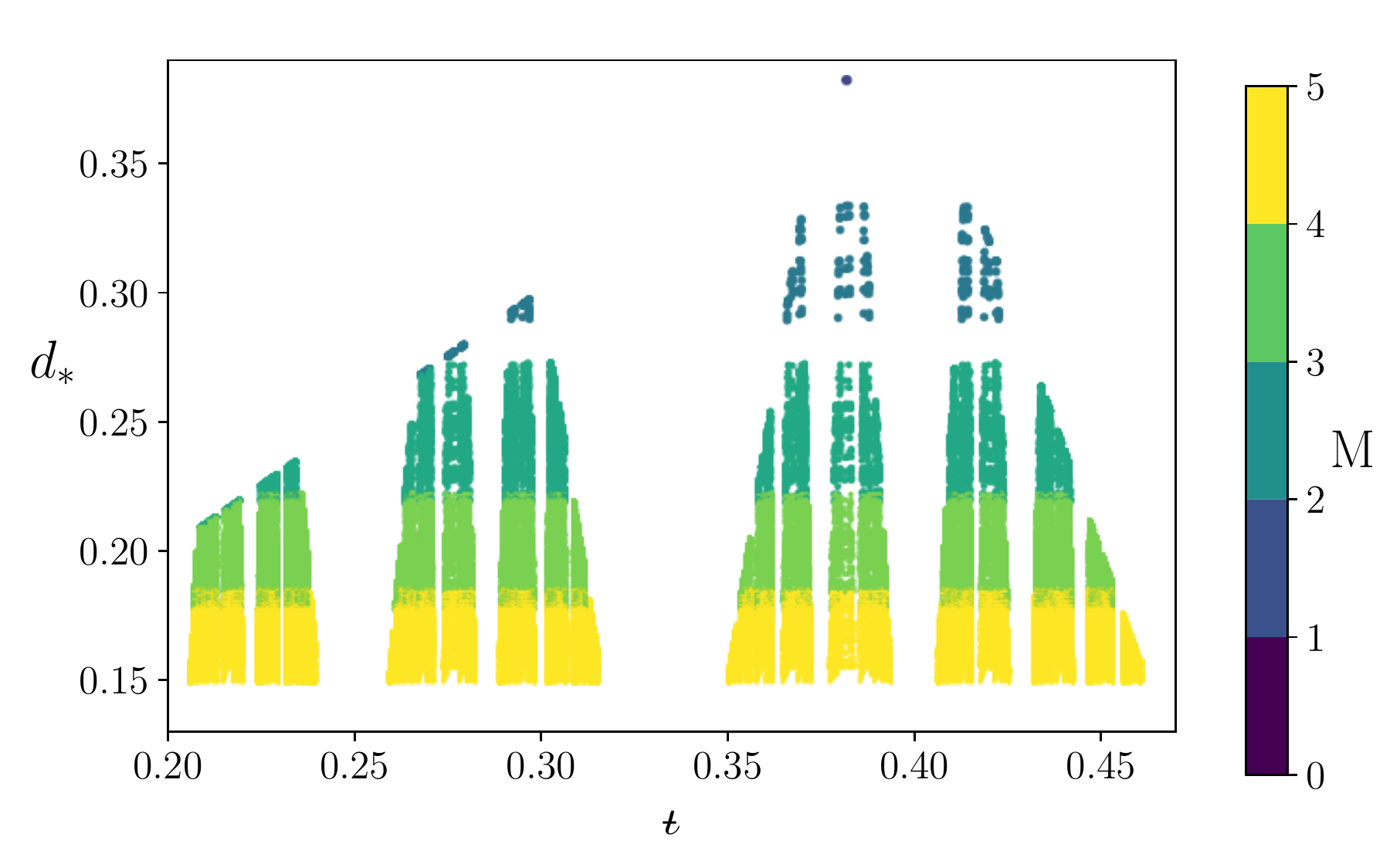}
\caption{\label{fig:dCritMaximumElement}A database of noble irrational numbers $\omega$ which have maximum element $M \le 5$ ($M$ is represented by color). Each $\omega$ has its location on the real line plotted against the value $d_*(\omega, k = 2)$. The real line $[0,1]$ is only half shown, since these calculations are symmetric about $\iotabar = \sfrac{1}{2}$.}
\end{figure}

\begin{itemize}
  \item As $k$ decreases, the measure of $\mathcal{D}_{d,k}$ decreases monotonically.
  \item As a rule, irrationals with maximum element $M$ are more robust than those with maximum element $N$ if $M < N$. This strict condition can be seen in the clean horizontal divisions between irrationals with different maximum elements, shown in Fig.~\ref{fig:dCritMaximumElement}.
  \item The parameter $d_*$ clarifies how robust an irrational surface is to perturbation, just as Greene's residue $f$ does.
  \item The most noble surface in our system (\mbox{$\iotabar = 1/\varphi$} or \mbox{$1 - 1/\varphi$}) is the most robust. This can be seen in the uppermost point of Fig.~\ref{fig:dCritMaximumElement}.
  \item Just as Greene identifies KAM surfaces wherever $f < 1$, the equilibria here have pressure gradients only on surfaces with Diophantine rotational-transform.
  \item Greene's trivial case of connected stochasticity occurs for the standard map parameter $ k > k_c^*$, whereas the pressure profile here is trivially zero for Diophantine parameters $k \le 2$ or $d^* \ge 2 - \varphi$.
\end{itemize}

Despite these gains in understanding, a more precise model that predicts the robustness of a given irrational without calculating $d_*$ has not yet been derived analytically. Various rules that dictate the robustness of a subset of irrationals are rarely valid when applied to a larger range.

\section{\label{section:profiles}Fractal profiles in ideal MHD}
 
Armed with a numerical discretization method that converges to the well-posed mathematics of a fully fractal pressure, we can now calculate the plasma profiles compatible with a Diophantine pressure gradient. As described in the introduction, toroidal geometry necessitates a non-smooth pressure profile due to the small-denominator resonances between toroidal and poloidal field line rotation. However, solving for ideal profiles in a fully 3D system introduces many difficulties \cite{Weitzner, Boozer}. Calculating the fractal pressure in cylindrical geometry provides a test-bed for isolating the effects of fractal pressure profiles before taking on the full 3D case.
 
The fundamental force-balance equation of ideal MHD equilibria is $\nabla p = \vect{J} \times \vect{B}$. Given Ampere's law \mbox{$\vect{J} = \nabla \times \vect{B}$} (setting $\mu_0 = 1$), the axial magnetic field equation becomes
\begin{equation} B_z' = -\frac{1}{R^2 + \iotabar^2 r^2} \bigg( \frac{p' R^2}{B_z} + B_z( r^2 \iotabar \iotabar' + 2r \iotabar^2)\bigg),\end{equation}
where $R$ is the major radius and all primes denote differentiation with respect to $r$. With $B_z'(r)$ determined as a function of $B_z(r),\ \iotabar(r)$, and $\nabla p(r)$, the magnetic field component $B_z(r)$ can be found at all points on the irregular grid for $r$. 

\begin{figure}
\centering
\includegraphics[width=0.5\textwidth]{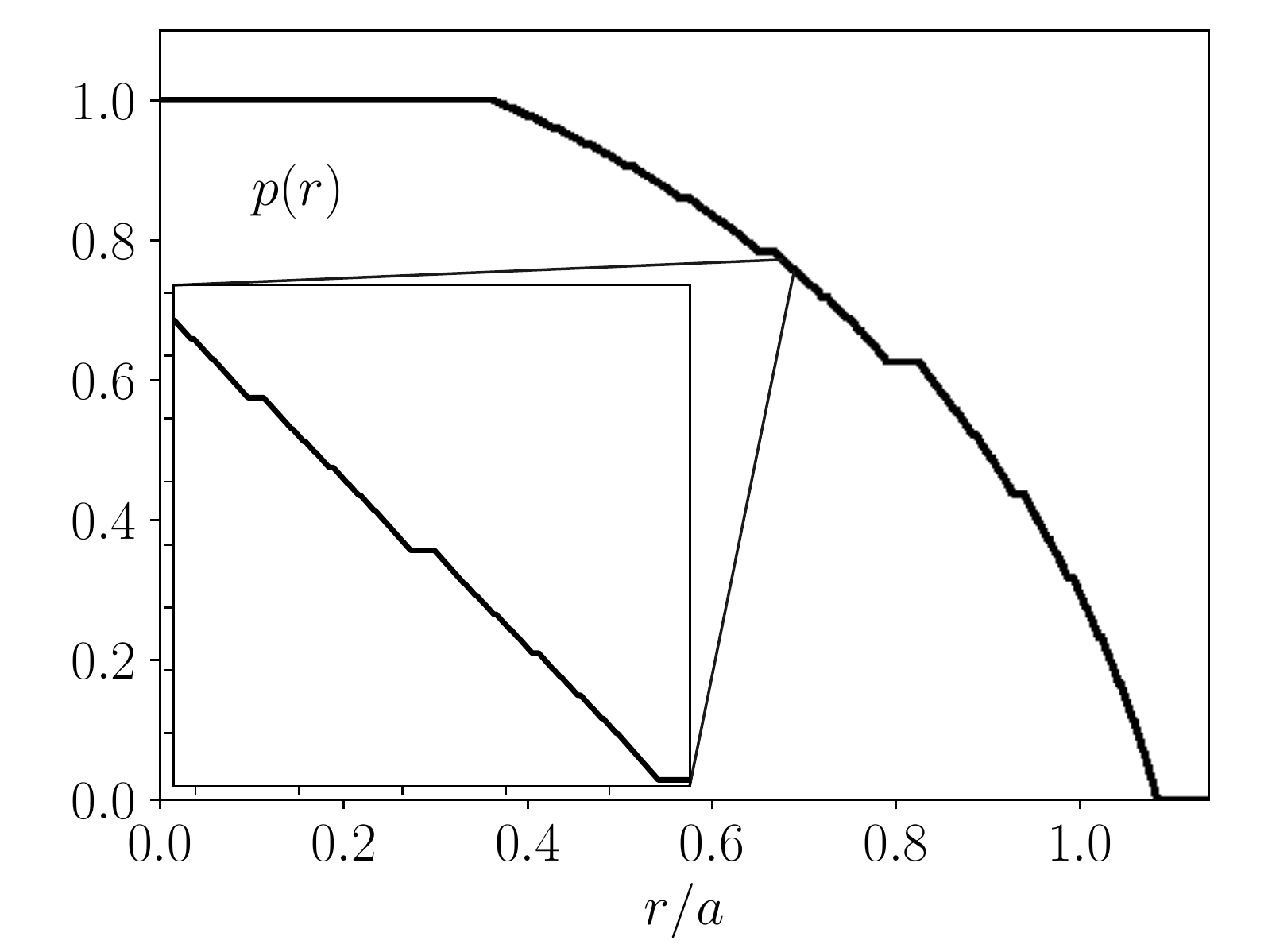}
\caption{\label{fig:pProfile}The pressure as a function of normalized minor radius. All calculations use a grid resolution with $\mathcal{F}_{15}$, and Diophantine parameters $d = 0.1$ and $k = 2.2$.}
\end{figure}

Only the free parameter $\iotabar(r)$ must be prescribed, here chosen as $\iotabar(r) = 1 - 7r^2/8$ to correspond qualitatively with standard tokamak field structure. For convenience, the pressure gradient is simply chosen to be $-1$ on all surfaces where $\iotabar$ is Diophantine; the pressure gradient $p'(\iotabar)$ could be multiplied by any smooth function to reflect various physical systems, without changing its fractal structure. For these calculations, $p'(\iotabar)$ was calculated by flattening rationals on the Farey tree $\mathcal{F}_{15}$, using Diophantine parameters $d = 0.1$ and $k = 2.2$. This profile was then translated to $p'(r)$ via the prescribed rotational-transform profile above. The resulting pressure profile is shown in Fig.~\ref{fig:pProfile}.

\begin{figure}
\centering
\includegraphics[width=0.5\textwidth]{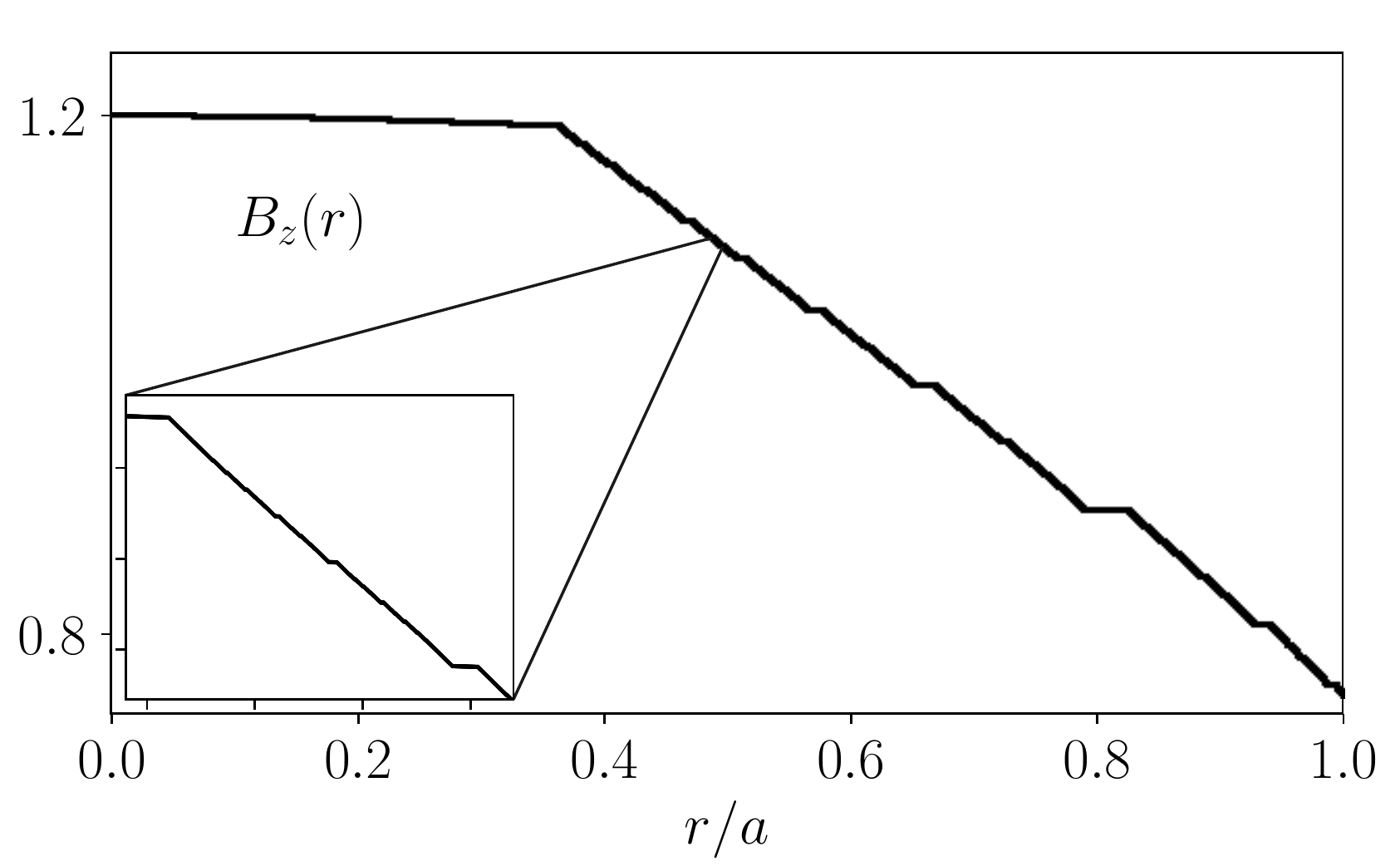}
\caption{\label{fig:Bprofile}The axial component of the magnetic field, $B_z(r)$.}
\end{figure}

This pressure gradient profile $p(r)$ can be input into the cylindrical differential equation above, and finally solved for $B'_z(r)$ with a piecewise fourth-order Runge-Kutta algorithm. Then, the remaining field and current components $B_\theta(r),\ J_\perp(r)$, and $J_\parallel(r)$ can all be found by simple multiplication and differentiation of the result for $B_z(r)$. The magnetic field structure, shown in Fig.~\ref{fig:Bprofile}, parallels the pressure: it is continuous, but only smooth on rational intervals. 

\begin{figure}[h]
\centering
\includegraphics[width=0.5\textwidth, bb=0 0 475 318]{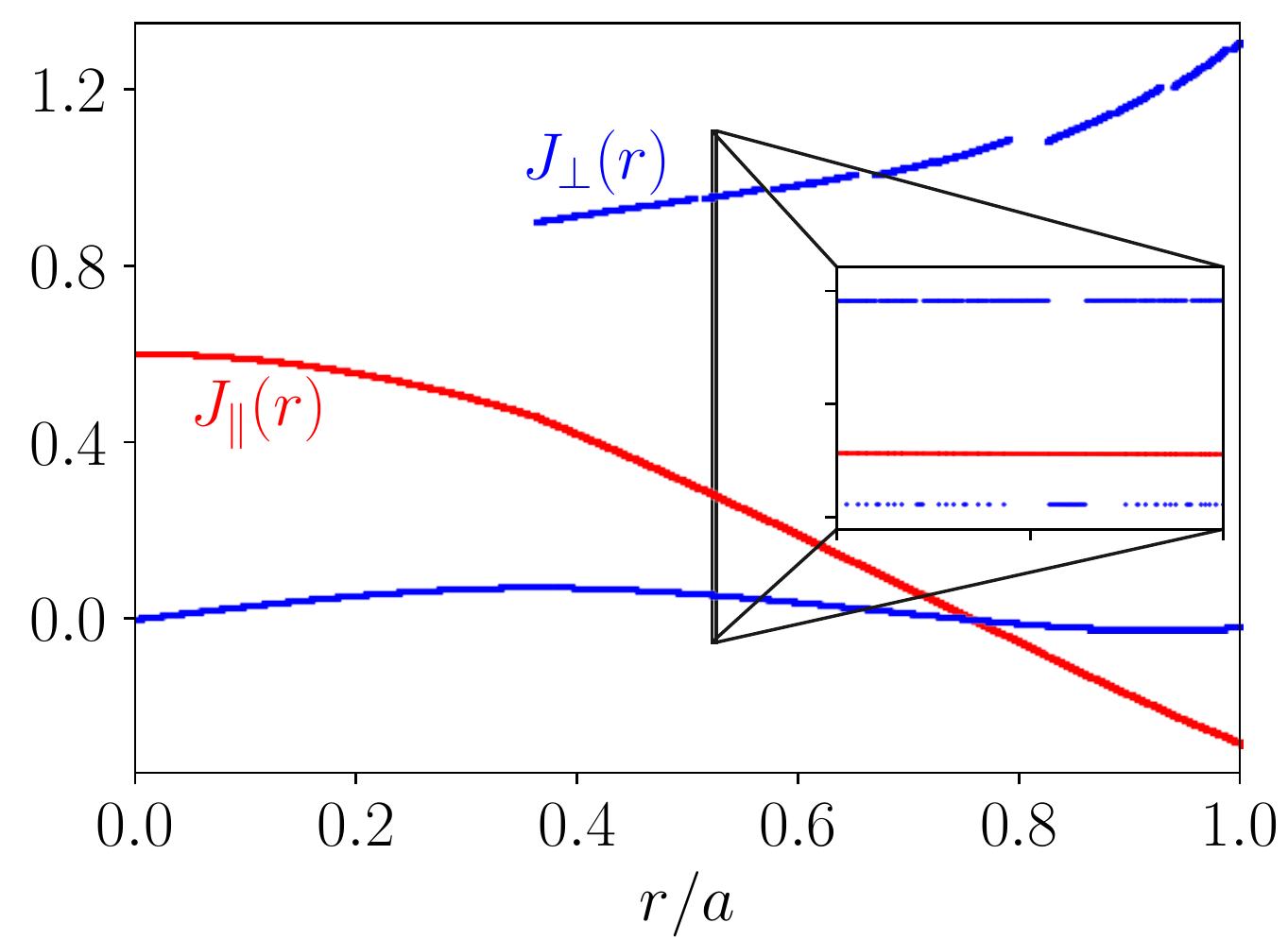}
\caption{\label{fig:currentProfile}Two components of the plasma current $\vect{J}$, shown as a function of normalized minor radius.}
\end{figure}

As shown in Fig.~\ref{fig:currentProfile}, \change{the perpendicular current reflects} the structure of the pressure gradient: the current is discontinuous on all Diophantine numbers. In the discretized case shown here, the discontinuities occur between rational and irrational intervals, but in the fractal limit there are no irrational intervals: each of the discontinuities is pointwise in the infinite-resolution case. \change{Interestingly, this discontinuous behavior is only present in the perpendicular current $J_\perp(r)$.}

Non-smooth pressure has introduced a fundamentally different structure into the otherwise straightforward cylindrical equilibrium. The pressure and magnetic field may be continuous, but the current density jumps between two distinct solutions depending on whether the rotational-transform is Diophantine or not. Such radically strange behavior, with interleaved sets of flux surfaces maintaining disjoint current densities, must be a unique product of ideal MHD, since non-ideal effects like finite Larmor radii or resistivity will smooth out the equilibrium. What will happen to these current density discontinuities when the pressure is smoothed?

\begin{figure}
\centering
\includegraphics[width=0.5\textwidth, bb=0 0 420 453]{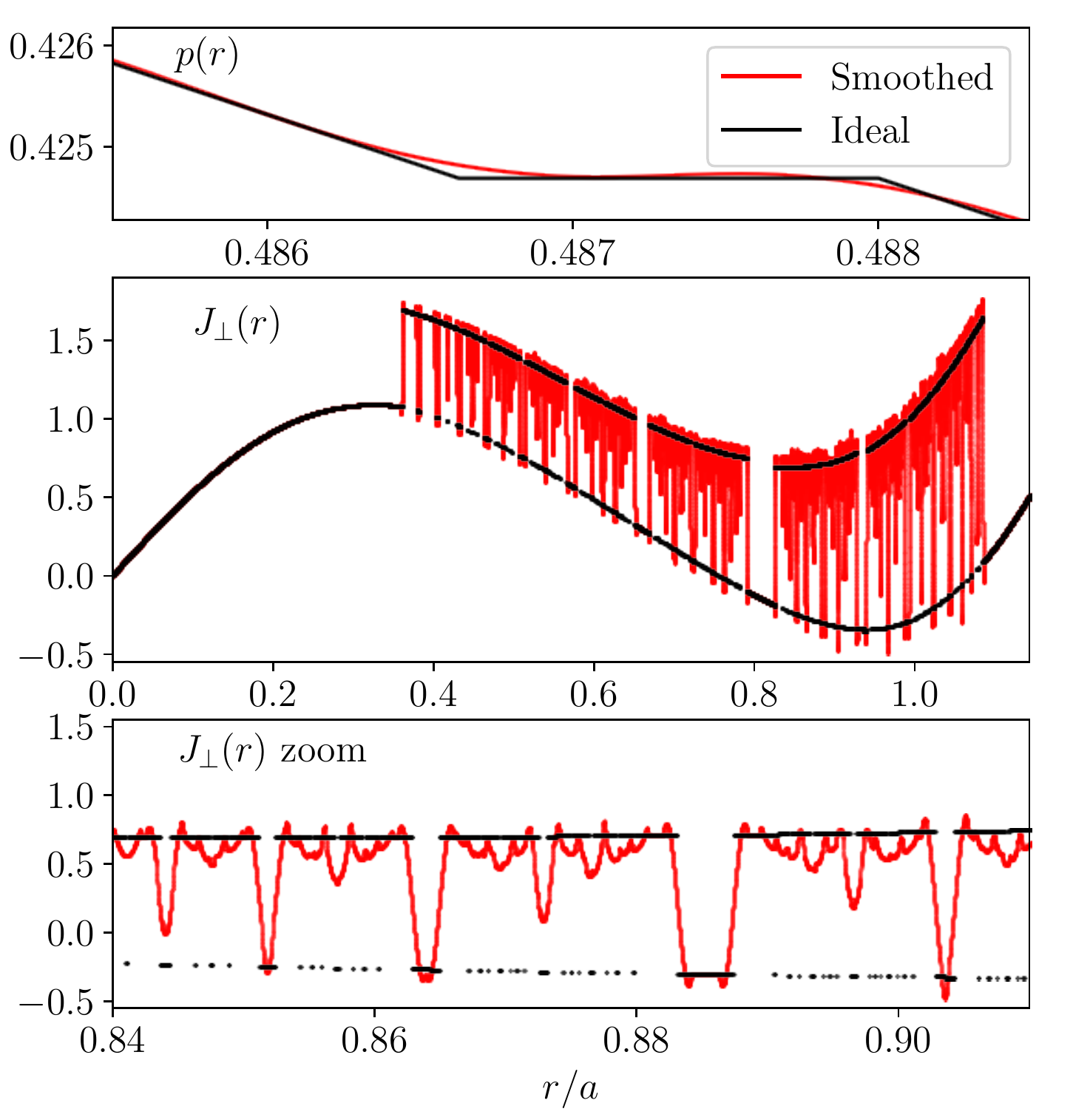}
\caption{\label{fig:smoothedCurrentProfile}Shown is a smoothed equilibrium which rounds the sharp corners of the input fractal pressure profile. On top, a zoom of the pressure profile is shown around a single flattened region; in the center, the full range of $J_\perp(r)$ is shown; on the bottom, a zoom into $J_\perp(r)$ displays the oscillatory structure of the smoothed current density.}
\end{figure}

As a proxy for introducing resistive effects, we repeat the same cylindrical equilibrium equation as above, but instead consider a smoothed version of the fractal pressure profile. To eliminate non-smooth corners between rational and irrational regions, the pressure profile was spatially smoothed with a third-order Savitzky-Golay filter \cite{Savitzky}. This filter leads to slight overshoots, but has sufficient response to round corners but maintain flat pressure on the smallest rational regions. The smoothed pressure was fed through the same differential equation solver for $B_z(r)$ as before to compute the resulting ``smooth'' MHD equilibrium. Results for the smoothed $J_\theta(r)$ are shown in Fig.~\ref{fig:smoothedCurrentProfile}. The current density now oscillates continuously between the rational and irrational values from the ideal case. Where the measure of Diophantine $\iotabar$ is high, $J_\theta$ remains mostly on the upper irrational curve. It dips toward lower values only when the measure of rational regions is significant. As is seen in the figure, this behavior appears stochastic from afar, but ultimately reflects the underlying order of the Diophantine numbers.

While the fractally flattened pressure suppresses infinite currents on rational flux surfaces, it results in discontinuities in $\vect{J}$ on all sufficiently irrational surfaces. The smoothed variant shown above replaces these discontinuities with oscillations which are just as unusual. These fractal profiles are physically acceptable, whereas infinite currents from small denominators were not. Still, they diverge from standard assumptions about smooth variations of plasma parameters. Ideal MHD and many other descriptions of confined plasmas predict this type of fractal behavior because non-smoothness is an inescapable consequence of 3D resonances.

Following the discussions on grid mechanics in Section \ref{section:grid}, we have confidence that these numerical representations of plasma equilibria are correctly converging on a true fractal limit. The explorations into the measure of the Diophantine set in Section \ref{section:robust} further support that these equilibria rest on sturdy mathematical ground. The Diophantine pressure profile we defined in Section \ref{section:Diophantine} is a toy model that oversimplifies the full toroidal case, but it still captures the difference between a confining plasma where pressure gradients are possible and one that is dominated by flattening rational resonances. 

Identifying the distribution of rational and irrational rotational-transform in a given equilibrium provides insight into different modes of plasma behavior. The measure-theoretical quantity $\mu(\mathcal{D}_{d,k})$ is a quantitative measure of the plasma volume fraction where currents, pressures and field behave in qualitatively ``irrational'' ways. Any prediction about this fractal behavior relies not just on numerical methods for calculating profiles, but on fundamental number and measure theory to ensure that numerics converge appropriately.

\begin{acknowledgements}
This work was supported by the U.S. Department of Energy and its Office of Science and Fusion Energy Sciences through contract DE-AC02-09CH11466.
\end{acknowledgements}

\end{document}